# A Probabilistic Approach to Power System State Estimation using a Linear Algorithm

Martin R. Wagner, Marko Jereminov, Amritanshu Pandey, Larry Pileggi
Department of Electrical and Computer Engineering
Carnegie Mellon University
Pittsburgh, PA, USA

*Abstract*— An equivalent circuit formulation for power system analysis was demonstrated to improve robustness of Power Flow and enable more generalized modeling, including that for RTUs (Remote Terminal Units) and PMUs (Phasor Measurement Units). These measurement device models, together with an adjoint circuit based optimization framework, enable an alternative formulation to Power System State Estimation (SE) that can be solved within the equivalent circuit formulation. In this paper, we utilize a linear RTU model to create a fully linear SE algorithm that includes PMU and RTU measurements to enable a probabilistic approach to SE. Results demonstrate that this is a practical approach that is well suited for real-world applications.

Power System State Estimation (SE) plays a crucial role in the Power Systems operations tool-chain. It uses measurement data and a Power Grid topology model to estimate the most likely state of a system. The original formulation of this widely studied problem was introduced by Schweppe [1], formulating it as a nonlinear weighted least squares optimization algorithm. Therein, the power system state is defined as a set of complex bus voltages in polar coordinates, leading to a highly nonlinear formulation that is known to be challenging to solve. In fact, even modern state estimators suffer from convergence issues in real-world scenarios [2].

SE algorithms traditionally consider data from Remote Terminal Units (RTUs), which are deployed as part of Power Systems' Supervisory Control and Data Acquisition (SCADA) systems. A second type of measurement device, the Phasor Measurement Unit (PMU) measures synchronized voltage and current phasors with high precision using geolocation systems. Algorithms exclusively considering PMU measurements are linear if they are formulated in cartesian coordinates [3], [4]. These linear SEs have been deployed to modern PS operations [3]. However, their success is limited by the amount of installed PMUs in today's systems [5]. To make use of both types of measurement devices hybrid SE approaches were developed [6], [7].

Alternative approaches to Schweppe's SE formulation exist. Notably, direct noniterative approaches considering SCADA data were recently proposed in [8] and [9]. Additionally, it has been shown that a cartesian coordinate formulation can lead to improved convergence [10]. A SE formulation based on the Current Injection Method (CIM) for Power Flow (PF) is proposed in [11].

The CIM was originally proposed as a formulation to solve the PF problem [12]. Building on it, the equivalent circuit formulation (ECF) was shown to enable robust PF convergence using circuit simulation techniques [13]. In addition, the ECF can incorporate any physics based model, including that from measurement devices such as PMUs and RTUs [14]. Using this modeling approach and an optimization toolbox, a linear state estimation algorithm was introduced in [15]. Separately, it was demonstrated that it is possible to extend the ECF to natively solve optimization problems, including SE, within the same framework [16]–[18].

Deployed in EMS systems, state estimators operate under hard time constraints. Since numerical approaches to probabilistic analysis are computationally intensive, only a few probabilistic SE approaches for special applications have been proposed [19],[20]. However, probabilistic approaches to general PS analysis have many advantages. According to [19], advantages of a probabilistic SE approach are: reduced standby resources, better ability to quantify voltage variabilities with intermittent renewable generation. [19] also discusses the possibility of probability-based market products. Makarov discusses advantages of a probabilistic PS planning and operations in [21] and further argues that not implementing a probabilistic approach would lead to "increasing risk of system failures, blackouts and near-misses", and "less economical system operation while addressing unexpected situations", among others. This suggests that a probabilistic SE could play a major part in comprehensive probabilistic PS analyses. For example, probabilistic SE naturally leads to probabilistic contingency analysis improving risk-awareness.

In this paper, we propose an algorithm for probabilistic SE. We use the modeling capabilities of the ECF to include a previously discussed PMU [14] and an RTU model that is mathematically equivalent to a recently introduced linear RTU model [15], thus defining a fully linear optimization-based SE algorithm that is solved completely within the ECF framework. Finally, we leverage the low computational cost and guaranteed solution of this algorithm to create a Monte Carlo based probabilistic SE.

Our results include a comparison of the deterministic part of our SE algorithm to the standard SE approach. Further, we demonstrate the unique features of a probabilistic SE by incorporating network uncertainties into the analysis. Finally, we discuss the scalability of this algorithm to real-world scenarios.

This work was supported in part by the Defense Advanced Research Projects Agency (DARPA) under award no. FA8750-17-1-0059 for the RADICS program, and the National Science Foundation (NSF) under contract no. ECCS-1800812.

## I. BACKGROUND

### A. AC Power System State-Estimation

Most state-of-the-art AC state estimators (ACSE) include measurements from RTUs, which sample voltage, current, and power magnitudes. Measurements from PMUs, which sample at higher rates and capture relative angle information are generally not included. As already mentioned, the ACSE can be formulated as a weighted least squares (WLS) algorithm with the objective function:

$$\min_x \ (z - h(x))^T W(z - h(x)) \tag{1}$$

where $z$ is the vector of RTU measurements that includes real and reactive power injection at nodes, voltage magnitude at the bus nodes and real and reactive power flows across the branches, $x$ is a vector of voltage state variables, $h$ is a vector valued measurement function describing the nonlinear relationship between estimated state and measurements, and $W$ is a diagonal weight matrix, where each term corresponds to the inverse of the variance of a given measurement $(1/\sigma_z^2)$. Notably, simplified decoupled linear versions of this algorithm exist [22]. However, this paper is concerned with an alternative algorithm to ACSE including PMU measurements.

### B. Equivalent Circuit Formulation

In the equivalent circuit formulation (ECF) for Power Flow, power systems are modeled in terms of current and voltage state variables. The ECF is formulated in cartesian coordinates using KCL equations on each node. Contrary to the classic PQV formulation for PF with a nonlinear network model and linear equations for constant power models, formulating PF in the ECF results in a fully linear PS network model, including linear models for transmission lines, transformers, and phase shifting transformers [13]. However, nonlinearities are now found in constant power models like the PQ-load model or the PV-generator model. The generic expression of a PF problem formulated with the ECF can be written in matrix form as

$$Y_{GB} V + I(V) = 0, \tag{2}$$

where $Y_{GB}$ is a linear bus admittance matrix, $V$ is the voltage state vector and $I(V)$ is a nonlinear function in terms of the state variables that contains the nonlinear PF models.

To numerically find a solution to this nonlinear problem the Newton-Raphson (NR) algorithm is used. The ECF uses the circuit formalism to solve the nonlinear problem. More precisely, nonlinear models are iteratively linearized and mapped into equivalent circuit elements, enabling the use of circuit simulation techniques to solve otherwise hard-to-solve problems [23].

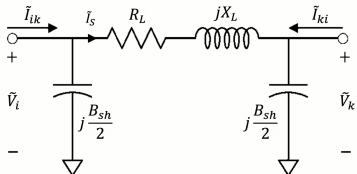

Fig. 1 Transmission line pi-model.

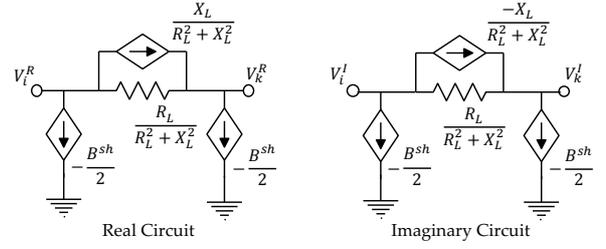

Fig. 2 ECF representation of a transmission line pi-model.

Fig. 1 and Fig. 2 show the mapping of a transmission line pi-model (Fig. 1) to its representation within the ECF (Fig. 2). We derive the series elements of this model by formulating Ohm's law between buses $i$ and $k$:

$$\tilde{I}_s = \frac{\tilde{V}_{ik}}{R_L + jX_L} \tag{3}$$

where $\tilde{V}_{ik}$ is the complex voltage difference between the two buses and $\tilde{I}_s$ is the complex current flowing through the series elements $R_L$ and $X_L$. Eq. (3) is further split into real and imaginary parts and formulated it in terms of currents

$$I_S^R = V_{ik}^R \frac{R_L}{R_L^2 + X_L^2} + V_{ik}^I \frac{X_L}{R_L^2 + X_L^2} \tag{4}$$

$$I_S^I = V_{ik}^I \frac{R_L}{R_L^2 + X_L^2} - V_{ik}^R \frac{X_L}{R_L^2 + X_L^2} \tag{5}$$

To map equations (4) and (5) to equivalent circuits, terms relating a voltage to its own current are interpreted as conductances, whereas terms that relate a current to a different voltage are represented by voltage-controlled current sources. Shunt elements of this model are derived in the same way. Finally, the equivalent circuit elements of each model are translated into matrix entries and included in $Y_{GB}$ of Eq. (2).

Nonlinear models are linearized and further translated into equivalent circuit elements using similar steps. The main difference is that contributions of nonlinear elements are updated for each NR iteration until convergence. Since nonlinear models are not included in the proposed SE formulation, we refer to [13] for a detailed discussion of handling nonlinear elements within the ECF and achieving robust convergence using circuit simulation techniques.

### C. Circuit Theoretic Optimization

Interestingly, it is possible to extend the ECF to solve optimization problems fully within the formulation. This was demonstrated for a PF feasibility optimization in [17], for AC Optimal Power Flow in [16], and SE in [18]. Because of its applicability for many optimization problems in power systems and beyond it was named equivalent circuit programming (ECP). ECP bases on adjoint-network theory [24]. Importantly, adjoint network-based approaches have been proposed in the power systems domain before [25], [26].

## II. STATE ESTIMATION IN THE ECF

AC Power System State Estimation (SE) can be formulated as an ECP optimization problem. This is done by first defining ECF measurement models for PMUs and RTUs that include variables reflecting measurement error. After integrating these models into a PS topology model, an optimization algorithm minimizing these variables can be defined. Notably, the PS topology model is the same that is used for PF, including branches, transformers, and shunts.

## A. PMU Model

A PMU measurement model consists of synchronized real and imaginary voltage measurements on a bus, as well as real and imaginary current measurements for branches from that bus. Both types of measurements are combined in the PMU model using current source conductances, as seen in Fig. 3.

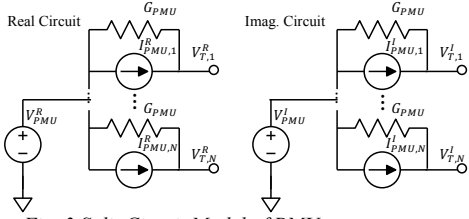

Fig. 3 Split Circuit Model of PMU measurements.

The currents through the source conductances $G_{PMU}$ are a measure of measurement error. This can be seen by assuming measurements that perfectly depict the state of the system. For this set of measurements, the voltages at the PMU model's terminals $V_{T,1..N}^{R,I}$ would be equal to the measured voltages $V_{PMU}^{R,I}$ resulting in zero currents through the source conductances $G_{PMU}$. Any measurement error would change this equilibrium and induce currents flowing through the source conductances. Hence, an algorithm to minimize measurement errors can be designed by minimizing the currents through these source conductances.

## B. Linear RTU Model

Remote Terminal Units (RTUs) measure current and voltage magnitudes and the power factor angles between current and voltages. These measurements are then converted into values of voltage magnitude and real and reactive powers [27]. This conversion simplifies including measurement data into the classic SE formulation that is formulated in terms of powers [1]. However, the original measurements lend themselves better for formulations using currents and voltages as state variables.

Our RTU model condenses multiple branch-current measurements into one net bus-injection current measurement [15]. While some amount of measurement information is lost with this compression, we gain a fully linear RTU model that enables our probabilistic SE algorithm. We believe that the resulting probabilistic system state and its associated information gain makes up for this initial information loss. Notably, a post-processing step can extract single current measurement quality information or give information about a branch model's quality.

The model maps the condensed measurements into an equivalent admittance at the measurement voltage.

$$Y_m = \frac{S_{RTU}}{V_M^2} = \frac{I_M}{V_M}[\cos(\phi_M) + j\sin(\phi_M)] \quad (6)$$

Here, $V_M$ is the measured bus voltage magnitude. Its equivalent admittance $Y_m = (G_M + jB_M)$ can be found by either using the converted power measurement $S_{RTU}: (P_{RTU} + jQ_{RTU})$, or the condensed original current measurement $I_M$ and the power factor angle measurement $\phi_M$. To complete the RTU model, the measurement admittance $Y_m$ is further augmented by current sources. Non-zero values of these current sources indicate measurement-error. Hence, the source contributions are later minimized in the SE optimization algorithm. Notably, this is similar to the earlier mentioned feasibility PF algorithm [17]. The equivalent circuit model of the RTU is depicted in Fig. 4.

A mathematically equivalent linear RTU model was recently proposed in [15], where RTU measurements are represented by current sources. Therein, an optimization problem is further defined and solved using general purpose optimization solvers. This paper proposes an algorithm that is fully solved within the ECF.

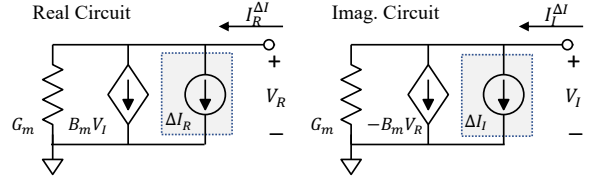

Fig. 4 Equivalent circuit of the linear RTU model.

## C. Defining the SE optimization problem

We discussed how both measurement models include a metric for measurement error. Based on these metrics, we can now define a minimization-based SE algorithm

$$\min \mathcal{F}_e = \frac{1}{2}\left\|I_{G_{PMU}}^{R,I}(V)\right\|_2^2 + \frac{1}{2}\left\|I_{RTU}^{R,I}\right\|_2^2 \quad (7)$$

that is subject to network constraints defined by the power system topology:

$$I_i^R + \sum_{k=1}^{n_B}[G_i(V_i^R - V_k^R) - B_i(V_i^I - V_k^I)] = 0 \quad (8)$$

$$I_i^I + \sum_{k=1}^{n_B}[G_i(V_i^I - V_k^I) + B_i(V_i^R - V_k^R)] = 0 \quad (9)$$

Here, $\tilde{V}_i = V_i^R + jV_i^I$ and $\tilde{V}_k = V_k^R + jV_k^I$ are the complex bus voltages at bus $i$ and $k$, respectively, $G_i$ is the bus admittance, $B_i$ the bus susceptance, and the summation is over each of the $n_B$ branches connected to bus $i$. The bus injection $\tilde{I} = I_i^R + jI_i^I$ further depends on the type of measurement device on the bus. An RTU's contribution to it is:

$$I_i^R = G_m V_i^R - B_m V_i^I + I_{RTU}^R \quad (10)$$

$$I_i^I = G_m V_i^I + B_m V_i^R + I_{RTU}^I \quad (11)$$

where $Y_m = G_m + jB_m$ is the RTU's measurement admittance defined in (6), and $I_{RTU}^{R,I}$ are the currents to be minimized. For a PMU measured bus we assume a single current injection measurement without loss of generality and arrive at (compare Fig. 3):

$$I_i^{R,I} = I_{PMU}^{R,I} + I_{G_{PMU}}^{R,I} \quad (12)$$

Importantly, $I_{G_{PMU}}^{R,I}$, the current through the PMU's source conductance $G_{PMU}$, is a dependent variable and can be expressed by formulating Ohm's law

$$I_{G_{PMU}}^{R,I} = G_{PMU}(V_i^{R,I} - V_{PMU}^{R,I}) \quad (13)$$

where $V_{PMU}^{R,I}$ are the real and imaginary voltage measurements. Notably, all constraints of this optimization (8)-(13) are linear. This problem can be solved using any optimization toolbox.

## D. Probabilistic Linear State Estimation

In this paper, we are solving this SE optimization problem as an ECP problem within the ECF framework. For this, we represent the network constraints by an aggregation of primal and dual network models (see [17] for details) as well as primal and dual measurement models. To obtain the best estimate of the network state, we minimize the norm of the currents ($I_{G_{PMU}}^{R,I}$ and $I_{RTU}^{R,I}$) that correspond to the measurement errors. Since all models exclusively comprise of linear circuit elements and the objective function is quadratic, this formulation results in a *fully linear State Estimation algorithm*.

This linear algorithm enables a probabilistic approach to SE that is able to capture additional uncertainties prevalent in power system models. Mainly, these are inherent uncertainties in the system topology model, in addition to the noise in grid measurements. Traditional state estimation approaches are generally deterministic and do not consider this variability. Hence, the results of a state estimation do not reflect the full possible distribution of states, leaving the grid operator less informed about the system. A probabilistic analysis of the estimated system state, including both measurement noise and network uncertainties, enables a much better understanding of possible grid states.

To implement this stochastic SE, we extended our prototype tool SUGAR (Simulation with Unified Grid Analysis and Renewables) that has previously been shown to have a robust framework for Monte Carlo based probabilistic analysis [28] with RTU and PMU measurement models. This SUGAR C++ implementation features a thread-level parallel Simple Random Sampling Monte Carlo engine that uses a fast modern pseudo random number generator with excellent statistical properties [29]. A hierarchical algorithm efficiently creates and evaluates Monte Carlo samples in parallel. Uncertainties for our SE algorithm are integrated by re-introducing measurement uncertainties as stochastic variables into our measurement models. Additionally, any other model parameter can be specified as a stochastic variable.

## III. RESULTS

We generate synthetic measurement data to validate our SE formulation using the following approach: First, we solve a power flow base case. This power flow solution is interpreted as the true system state, with a vector of complex state-voltages denoted as $\bar{X}$. Then, based on statistical inputs, either PMU or RTU measurement models are randomly assigned to every bus of the system, thereby replacing the original power flow models on that bus. The power flow topology remains unchanged. Finally, we assign measurement values to each PMU and RTU by superimposing randomly created measurement errors on the pre-calculated true system states. In the following experiments, the assignment of a PMU or RTU model to a certain bus remains unchanged when multiple SE samples are created.

We examine the state estimation algorithms using the following measures:

$$x_\sigma = (\hat{X} - \bar{X})^T (\hat{X} - \bar{X}) \quad (14)$$
$$x_{max} = \max|\hat{X} - \bar{X}| \quad (15)$$
$$r_z = (z - h(\hat{X}))^T (z - h(\hat{X})) \quad (16)$$

here $x_\sigma$ is the sum of squared errors over the real and imaginary voltage state variables, $x_{max}$ is the maximum voltage vector deviation, and $r_\sigma$ is the sum of squared measurement residuals of the optimization.

## A. Comparison against Static AC – WLS State Estimator

First, we compare the deterministic ECF based SE algorithm with the Matpower implementation of a traditional AC WLS SE algorithm [30] that was adapted to include current injection measurements on each bus. As previously mentioned, phasor measurements are generally not part of this WLS formulation. However, our current formulation requires at least one phasor measurement to set the system's reference angle. To enable a valid comparison between the two algorithms, we select a zero-injection node to set the reference angle of the system. This eliminates the influence of the PMU's current phasor measurement for this comparison.

A second difference between the two formulations is how the weights are applied to different measurements. The WLS algorithm weighs every single measurement value by the matrix $W$ that scales its influence within the objective (1). Our formulation does not include every measurement in the objective function but maps them into models that are further used in the optimization. Hence, only weighting of measurement models as a whole is possible. In order to be conservative in this comparison, we weigh individual measurements within the WLS estimator by the inverse of their standard deviations, whereas within the ECF based algorithm we weigh each measurement model equally.

TABLE 1: ECF SE PERFORMANCE MEASURES FOR DIFFERENT SYSTEMS[1]

|    | $x_\sigma \pm ci_{99\%}$ | | $x_{max} \pm ci_{99\%}$ | | $r_z \pm ci_{99\%}$ | |
|----|---------|---------|---------|---------|---------|---------|
| #1 | 6.50e-3 | 8.00e-3 | 1.14e-2 | 1.25e-2 | 2.04e-2 | 2.29e-2 |
| #2 | 3.80e-3 | 4.60e-3 | 7.30e-3 | 8.00e-3 | 3.67e-2 | 3.95e-2 |
| #3 | 2.25e-2 | 2.76e-2 | 0.97e-2 | 1.04e-2 | 4.38e-1 | 4.72e-1 |
| #4 | 5.92e-2 | 7.16e-2 | 1.31e-2 | 1.43e-2 | 4.05e-1 | 4.49e-2 |
| #5 | 1.04e-1 | 1.26e-1 | 1.27e-2 | 1.41e-2 | 1.22    | 1.29    |

Table 1 shows performance measures of the ECF based SE algorithm for five openly available power system test cases [31], [32]. Synthetic sets of measurements were created using the methodology described earlier in this section with the following statistics: RTUs have 1% normally distributed real and reactive power injection measurement uncertainty, and 0.4% voltage magnitude measurement uncertainty. The single PMU that sets the reference angle within the ECF is assumed to be a perfect measurement. Its source conductance $G_{PMU}$ is set to 10 p.u. All uncertainties are based on the measures' "true states".

To perform a valid comparison between the two algorithms, we ran multiple SE samples until we reached an accuracy such that at least one performance measure ((14)-(16)) was within $\pm 5\%$ of its mean with 99% confidence. On average 544 instances were run for each case.

---

[1] #1: IEEE-118 bus system; #2: case_ACTIVSg500; #3:1888-bus RTE model; #4: case_ACTIVSg2000; #5: 6515-bus RTE model.

TABLE 2 WLS SE PERFORMANCE MEASURES FOR DIFFERENT SYSTEMS[1]

|    | $x_\sigma \pm ci_{99\%}$ | | $x_{max} \pm ci_{99\%}$ | | $r_z \pm ci_{99\%}$ | |
|----|----------|----------|----------|----------|----------|----------|
| #1 | 3.60e-3  | 4.60e-3  | 7.20e-3  | 8.00e-3  | 2.20e-3  | 2.30e-3  |
| #2 | 2.30e-3  | 2.90e-3  | 4.80e-3  | 5.40e-3  | 8.90e-3  | 9.00e-3  |
| #3 | 1.07e-2  | 1.33e-2  | 7.10e-3  | 7.60e-3  | 3.54e-2  | 3.57e-2  |
| #4 | 2.44e-2  | 3.06e-2  | 7.70e-3  | 8.50e-3  | 3.58e-2  | 3.61e-2  |
| #5 | 6.33e-2  | 7.92e-2  | 8.40e-3  | 9.30e-3  | 1.25     | 1.26     |

From Table 2 we see slightly better results for the nonlinear WLS SE formulation as compared to our linear ECF based State Estimator (Table 1). Notably, this is under conditions that are designed to compare a best-case scenario for the WLS, where weighting can be applied due to good knowledge of uncertainties, and possible divergence is avoided by good initial conditions.

Different to the nonlinear WLS approach, our formulation guarantees a solution due to its linearity and is superior to the WLS algorithm in terms of computational complexity. This enables a probabilistic approach to SE, resulting in a more complete (probabilistic) picture of a system's state.

*B. Probabilistic State Estimation*

We study our probabilistic algorithm on the 1888-bus RTE system model of the French transmission grid that is openly available [31]. This system size is a realistic abstraction of a regional size grid operator in the US. NYPA's EMS model for example comprises of around 1600 buses [33].

Our algorithm is tested in the following way: We synthetically create a single set of measurement data, where 10% of system buses are randomly selected to be PMU measured. Of these PMUs, 40% are assumed to be perfect measurements. The remaining 60% of PMUs have 0.02% uncertainty for all their measures (i.e. real and imaginary currents and voltages). All other buses are RTU measured with RTU uncertainties identical to the previous experiment. Additionally, we add network model uncertainties that are defined in Table 3.

TABLE 3 STANDARD DEVIATIONS OF NORMALLY DISTRIBUTED UNCERTAINTY VALUES OF TRANSMISSION LINE AND TRANSFORMER SERIES ELEMENTS

| Network uncertainties | $\sigma_R$ [% of mean] | $\sigma_X$ [% of mean] |
|---|---|---|
| Transmission line | 5% | 0.5% |
| Transformer | 1% | 0.1% |

Fig. 5 shows selected results from a probabilistic SE simulation with one million samples. The probabilistic density functions (PDFs) in Fig. 5 depict the voltage magnitude and angle of bus 1337 and the real and reactive branch flow on the branch that connects the buses 1337 and 311, which has one of the highest loadings in the system. The (PDFs) without network uncertainties are shown in green, whereas the PDFs with network uncertainties are shown in blue. It can be seen that the true system state $\bar{X}$, shown as a red-dotted line, is within the distributions for all PDFs. Moreover, we observe low influences of network uncertainties for voltage angle and real power in Fig. 5. However, standard deviations of the voltage magnitude and reactive power distributions were raised by 4.95%, and 21.3%, respectively.

Interestingly, the probabilistic algorithm allows us to make statements about probability of states. This is not possible in a deterministic setting. We find, for example that the probability of an absolute value of real power flow on the presented branch of greater than 15 p.u. is 0.244%. Alternatively, we find that there is less than a 1% chance for the real power flow on this branch to be higher than 14.73 p.u. (with a 99% confidence interval of $\pm$ 0.0023 p.u.). Statements of this sort are a valuable addition for system awareness and can lead to improved risk awareness, which is especially pertinent in modern grids with ever increasing uncertainties and reduced margins of error.

The algorithm to estimate confidence intervals of percentiles in MC solutions is based on order statistics and Binomial distribution [34]. Estimated confidence intervals are an important measure to characterize the quality of MC results. These algorithms can either be used to approximate the amount of necessary MC samples prior to running the simulation, or they can be calculated during the simulation to finalize the algorithm when the required accuracy is reached.

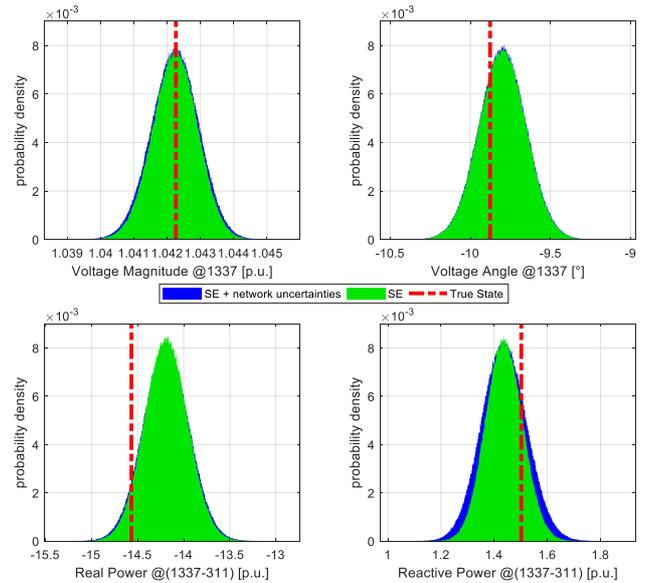

*Fig. 5 Selected probabilistic density functions (PDFs) of our SE algorithm for the 1888-bus RTE case, with (blue) and without (green) including network uncertainties, and the true state of the system.*

*C. Performance and Scaling of the Algorithm*

The linear probabilistic algorithm that was implemented in C++ was used to evaluate the $10^6$ Monte Carlo samples of the 1888-bus case in 7 minutes and 9 seconds on an Intel Xeon server CPUs (E5-2680) running on 2.40GHz using 45 of 56 possible threads. Monte Carlo samples of this algorithm are independent and have close to constant runtime, hence it is expected to scale linearly with additional CPUs. Additionally, it has been shown that similar (nonlinear) algorithms have single iteration runtimes of under one second for systems with close to $10^5$ buses [16]. This suggests that our algorithm has the potential to scale well enough to be used in an operations setting for up to continental grid interconnection sized systems on currently available hardware.

*D. Discussion and Paths forward*

A Simple Random Sampling Monte Carlo approach requires a big sample size to arrive at statistically valuable results. Methods to reduce this sample count while maintaining accuracy under certain assumptions exist. They commonly referred to as Quasi Monte Carlo methods. Alternatively, statistical theories like the Extreme Value Theory or the Large Deviation Theory can be applied to improve the statistical quality of statements of this kind. We aim to study applicability of both approaches in future.

Our current algorithm assumes independence of every probabilistic variable, which is generally not the case. Measurements that are only connected through a single network connection, or more general measurements that are in close proximity to each other, are likely to have some correlation with each other. Future work will consider adding models of these correlations to further improve the quality of the results.

## IV. Conclusion

This paper proposes a Monte Carlo based linear probabilistic State Estimation algorithm. It formulates SE as optimization problem that is expressed in terms of equivalent circuit models. A linear RTU model and a linear PMU model together with a linear topology model facilitate this fully linear algorithm without simplifications. We compare our approach with the traditional WLS formulation for SE for different systems. With similar estimation quality, our approach is found to be computationally superior and is provably convergent. Leveraging this attribute, we implement a probabilistic SE algorithm by reintroducing measurement uncertainties. To further demonstrate advantages of this probabilistic approach we add network model uncertainties as variables. Results suggest that our linear probabilistic SE algorithm is well suited for use under operational time-constraints for medium sized systems and may scale up to interconnection sized systems. Finally, possible future refinements such as Quasi-MC approaches, model correlations, and potentially useful statistical theories are discussed.